\documentclass[runningheads]{llncs}
\pdfoutput=1
\usepackage{pscproc2,graphicx}

\newcommand{\Oh}[1]
    {\ensuremath{\mathcal{O} \hspace{-.5ex} \left( {#1} \right)}}
\newcommand{\polylog}[1]
    {\ensuremath{\mathrm{polylog} \hspace{-.5ex} \left( {#1} \right)}}

\begin{document}

\title{Fast and Compact Prefix Codes}
\author{Travis Gagie\inst{1}
    \and Gonzalo Navarro\inst{2}
    \and Yakov Nekrich\inst{3}}
\institute{Research Group in Genome Informatics\\
    Bielefeld University\\
    \email{travis.gagie@gmail.com}\\\mbox{}\\
    \and Department of Computer Science\\
    University of Chile\\
    \email{gnavarro@dcc.uchile.cl}\\\mbox{}\\
    \and Department of Computer Science\\
    University of Bonn\\
    \email{yasha@cs.uni-bonn.de}}
\maketitle

\begin{abstract}
It is well-known that, given a probability distribution over $n$ characters,
in the worst case it takes \(\Theta (n \log n)\) bits to store a prefix code
with minimum expected codeword length.  However, in this paper we first show
that, for any $0<\epsilon<1/2$ with \(1 / \epsilon = \Oh{\polylog{n}}\), it takes $\Oh{n \log \log (1 / \epsilon)}$ bits to store a prefix code with expected codeword length within $\epsilon$ of the minimum.  We then show that, for any constant \(c > 1\), it takes $\Oh{n^{1 / c} \log n}$ bits to store a prefix code with expected codeword length at most $c$ times the minimum.  In both cases, our data structures allow us to encode and decode any character in $\Oh{1}$ time.
\end{abstract}

\section{Introduction} \label{sec:intro}

Compression is most important when space is in short supply, so popular compressors are usually heavily engineered to reduce their space usage.  Theory has lagged behind practice in this area, however, and there remain basic open questions about the space needed for even the simplest kinds of compression.  For example, while compression {\em with} prefix codes is familiar to any student of information theory, very little has been proven about compression {\em of} prefix codes.  Suppose we are given a probability distribution $P$ over an alphabet of $n$ characters.  Until fairly recently, the only general bounds known seem to have been, first, that it takes \(\Theta (n \log n)\) bits in the worst case to store a prefix code with minimum expected codeword length and, second, that it takes $\Oh{n}$ bits to store a prefix code with expected codeword length within 1 of the minimum.

In 1998 Adler and Maggs~\cite{AM01} showed it generally takes more than \((9 / 40) n^{1 / (20 c)} \log n\) bits to store a prefix code with expected codeword length at most \(c H (P)\), where \(H (P)\) is $P$'s entropy and a lower bound on the expected codeword length.  (In this paper we consider only binary codes, and by $\log$ we always mean $\log_2$.)
In 2006 Gagie~\cite{Gag06a,Gag06b} (see also~\cite{Gag08}) showed that, for
any constant \(c \geq 1\), it takes $\Oh{n^{1 / c} \log n}$ bits to store a
prefix code with expected codeword length at most \(c H (P) + 2\).  He also
showed his upper bound is nearly optimal because, for any positive constant
$\epsilon$, we cannot always store a prefix code with expected codeword length
at most \(c H (P) + o (\log n)\) in $\Oh{n^{1 / c - \epsilon}}$ bits.  Gagie
proved his upper bound by describing a data structure that stores a prefix
code with the prescribed expected codeword length in the prescribed space and
allows us to encode and decode any character in time at most proportional to
its codeword's length.  This data structure has three obvious defects: when
\(c = 1\), it is as big as a Huffman tree, whereas its redundancy guarantee
can be obtained with just $\Oh{n}$ bits \cite{GM59}; when \(H (P)\) is small, a possible additive increase of 2 in the expected codeword length may be prohibitive; and it is slower than the state of the art.

In this paper we answer several open questions related to efficient
representation of codes. First, in Section~\ref{sec:additive} we show that,
for any \(0 < \epsilon < 1 / 2\) with \(1 / \epsilon = \Oh{\polylog{n}}\), it takes $\Oh{n
\log \log (1 / \epsilon)}$ bits to store a prefix code with expected codeword
length within $\epsilon$ of the minimum.  Thus, if we can tolerate an additive
increase of, say, $0.01$ in the expected codeword length, then we can
store a prefix code using only $\Oh{1}$ bits per character. Second, in
Section~\ref{sec:multiplicative} we show that, for any constant \(c > 1\), it
takes $\Oh{n^{1 / c} \log n}$ bits to store a prefix code with expected
codeword length at most $c$ times the minimum, with no extra additive increase.
Thus, if we can tolerate a
multiplicative increase of, say, $2.01$ then we can store a prefix code in $\Oh{\sqrt{n}}$ bits.  In both cases, our data structures allow us to encode and decode any character in $\Oh{1}$ time.

\section{Related work} \label{sec:related}

A simple pointer-based implementation of a Huffman tree takes $\Oh{n \log n}$ bits, and it is not difficult to show this is an optimal upper bound for storing a prefix code with minimum expected codeword length.  For example, suppose we are given a permutation $\pi$ over $n$ characters.  Let $P$ be the probability distribution that assigns probability \(1 / 2^i\) to the \(\pi (i)\)th character, for \(1 \leq i < n\), and probability \(1 / 2^{n - 1}\) to the \(\pi (n)\)th character.  Since $P$ is dyadic, every prefix code with minimum expected codeword length assigns a codeword of length $i$ to the \(\pi (i)\)th character, for \(1 \leq i < n\), and a codeword of length \(n - 1\) to the \(\pi (n)\)th character.  Therefore, given any prefix code with minimum expected codeword length and a bit indicating whether \(\pi (n - 1) < \pi (n)\), we can find $\pi$.  Since there are $n!$ choices for $\pi$, in the worst case it takes \(\Omega (\log n!) = \Omega (n \log n)\) bits to store a prefix code with minimum expected codeword length.

Considering the argument above, it is natural to ask whether the same lower
bound holds for probability distributions that are not so skewed, and the
answer is no. A prefix code is {\em canonical}~\cite{SK64} if the first
codeword is a string of 0s and any other codeword can be obtained from its
predecessor by adding 1, viewing its predecessor as a binary number, and
appending some number of 0s. (See, e.g.,~\cite{MT97,Kle00} for more recent
work on canonical codes.) Given any prefix code, without changing the length
of the codeword assigned to any character, we can put the code into canonical
form by just exchanging left and right siblings in the code-tree. Moreover, we
can reassign the codewords such that, if a character is lexicographically the
$j$th with a codeword of length $\ell$, then it is assigned the $j$th
consecutive codeword of length $\ell$. It is clear that it is sufficient to
store the codeword length of each character to be able of reconstructing such a
code, and thus the code can be represented in $\Oh{n\log L}$ bits, where $L$
is the longest codeword.

The above gives us a finer upper bound. For example,
Katona and Nemetz~\cite{KN76} showed that, if a character has probability $p$,
then any Huffman code assigns it a codeword of length at most about \(\log (1
/ p) / \log \phi\), where \(\phi \approx 1.618\) is the golden ratio, and
thus $L$ is at most about \(1.44 \log (1 / p_{\min})\), where $p_{\min}$ is
the smallest probability in $P$. Alternatively, one can enforce a value for
$L$ and pay a price in terms of expected codeword length.
Milidi\'u and Laber \cite{ML01} showed how, for any \(L > \lceil \log n \rceil\), we can
build a prefix code with maximum codeword length at most $L$ and expected
codeword length within \(1 / \phi^{L - \lceil \log (n + \lceil \log n \rceil -
L) \rceil - 1}\) of the minimum.  Their algorithm works by building a Huffman
tree $T_1$; removing all the subtrees rooted at depth greater than $L$;
building a complete binary tree $T_2$ of height $h$ whose leaves are those removed from $T_1$; finding the node $v$ at depth \(L - h - 1\) in $T_1$ whose subtree $T_3$'s leaves are labelled by characters with minimum total probability (which they showed is at most \(1 / \phi^{L - \lceil \log (n + \lceil \log n \rceil - L) \rceil - 1}\)); and replacing $v$ by a new node whose subtrees are $T_2$ and $T_3$.

A simple upper bound for storing a prefix code with expected codeword length
within a constant of the minimum, follows from Gilbert and Moore's
proof~\cite{GM59} that we can build an alphabetic prefix code with expected
codeword length less than \(H (P) + 2\) and, thus, within 2 bits of the
minimum. Moreover, in an optimal alphabetic prefix code, the expected codeword
length is within 1 of the minimum~\cite{Nak91,She92}
which, in turn, is within 1 of the entropy $H(P)$.
In an alphabetic prefix code, the lexicographic order of the
codewords is the same as that of the characters, so we need store only the
code-tree and not the assignment of codewords to characters.  If we store the
code-tree in a succinct data structure due to Munro and Raman~\cite{MR01},
then it takes $\Oh{n}$ bits and encoding and decoding any character takes time
at most proportional to its codeword length. This can be improved to $\Oh{1}$
by using table lookup, but doing so may worsen the
space bound unless we also restrict the maximum codeword length, which may in
turn increase in the expected codeword length.

The code-tree of a canonical code can be stored in just $\Oh{L^2}$ bits: By
its definition, we can reconstruct the whole canonical tree given only
the first codeword of each length. Unfortunately, Gagie's lower
bound~\cite{Gag06b} suggests we generally cannot combine results concerning
canonical codes with those concerning alphabetic prefix codes.

Constant-time encoding and decoding using canonical code-trees is simple.
Notice that if two codewords have the same length, then the difference between
their ranks in the code is the same as the difference between the codewords
themselves, viewed as binary numbers.  Suppose we build an $\Oh{L^2}$-bit
array $A$ and a dictionary $D$ supporting predecessor queries, each storing
the first codeword of each length.  Given the length of a character's codeword
and its rank among codewords of the same length (henceforth called its
offset), we can find the actual codeword by retrieving the first codeword of
that length from $A$ and then, viewing that first codeword as a binary number,
adding the offset minus 1.  Given a binary string starting with a codeword, we
can find that codeword's length and offset by retrieving the string's
predecessor in $D$, which is the first codeword of the same length; truncating
the string to the same length in order to obtain the actual codeword; and subtracting the first codeword from the actual codeword, viewing both as binary numbers, to obtain the offset minus 1.  (If $D$ supports numeric predecessor queries instead of lexicographic predecessor queries, then we store the first codewords with enough 0s appended to each that they are all the same length, and store their original lengths as auxiliary information.)  Assuming it takes $\Oh{1}$ time to compute the length and offset of any character's codeword given that character's index in the alphabet, encoding any character takes $\Oh{1}$ time.  Assuming it takes $\Oh{1}$ time to compute any character's index in the alphabet given its codeword's length and offset, decoding takes within a constant factor of the time needed to perform a predecessor query on $D$.  For simplicity, in this paper we consider the number used to represent a character in the machine's memory to be that character's index in the alphabet, so finding the index is the same as finding the character itself.

In a recent paper on adaptive prefix coding, Gagie and Nekrich~\cite{GN09} (see also~\cite{KN??}) pointed out that if \(L = \Oh{w}\), where $w$ is the length of a machine word, then we can implement $D$ as an $\Oh{w^2}$-bit data structure due to Fredman and Willard~\cite{FW93} such that predecessor queries take $\Oh{1}$ time.  This seems a reasonable assumption since, for any string of length $m$ with \(\log m = \Oh{w}\), if $P$ is the probability distribution that assigns to each character probability proportional to its frequency in the string, then the smallest positive probability in $P$ is at least \(1 / m\); therefore, the maximum codeword length in either a Huffman code or a Shannon code for $P$ is $\Oh{w}$.  Gagie and Nekrich used $\Oh{n \log n}$-bit arrays to compute the length and offset of any character's codeword given that character's index in the alphabet, and vice versa, and thus achieved $\Oh{1}$ time for both encoding and decoding.

A technique we will use to obtain our result is the {\em wavelet tree} of
Grossi et al.~\cite{GGV03}, and more precisely the multiary variant due to
Ferragina et al.~\cite{FMMN07}. The
latter represents a sequence $S[1,n]$ over an alphabet $\Sigma$ of size
$\sigma$ such that the following operations can be
carried out in $\Oh{\frac{\log \sigma}{\log\log n}}$ time on the RAM model
with a computer word of length $\Omega(\log n)$: (1) Given $i$, retrieve
$S[i]$; (2) given $i$ and $c \in \Sigma$, compute $rank_c(S,i)$, the number of
occurrences of $c$ in $S[1,i]$; (3) given $j$ and $c \in \Sigma$, compute
$select_c(S,j)$, the position in $S$ of the $j$-th occurrence of $c$. The
wavelet tree requires $nH_0(S) + \Oh{\frac{n\log\log n}{\log_\sigma n}}$ bits
of space, where $H_0(S) \le n\log\sigma$ is the {\em empirical zero-order
entropy} of $S$, defined as $H_0(S) = H(\{n_c/n\}_{c \in\sigma})$, where $n_c$
is the number of occurrences of $c$ in $S$. Thus $nH_0(S)$ is a lower bound
to the output size of any zero-order compressor applied to $S$.
It will be useful to write
$H_0(S) = \sum_{c\in\sigma} \frac{n_c}{n} \log \frac{n}{n_c}$.

\section{Additive increase in expected codeword length} \label{sec:additive}

In this section we exchange a small additive penalty over the optimal prefix
code for a space-efficient representation of the encoding, which in addition
enables encode/decode operations in constant time.

It follows from Milidi\'u and Laber's bound \cite{ML01} that, for any
\(\epsilon > 0\), there is always a prefix code with maximum codeword length
\(L = \lceil \log n\rceil + \lceil \log (2 / \epsilon)\rceil \) and expected
codeword length within
\[
\frac{1}{\phi^{L - \lceil \log (n + \lceil \log n \rceil - L) \rceil - 1}}
~~\le~~ \frac{1}{\phi^{L-\lceil\log n\rceil-1}}
~~\le~~ \epsilon^{-\log \phi} ~~<~~ \epsilon
\]
of the minimum. The techniques described in the previous section give a way
to store such a code in $\Oh{L^2 + n\log L}$ bits, yet it is not immediate how
to do constant-time encoding and decoding. Alternatively, we can achieve
constant-time encoding and decoding using $\Oh{w^2 + n\log n}$ bits for the
code-tree.

To achieve constant encoding and decoding times without ruining the space,
we use multiary wavelet trees. We use a canonical code, and sort the characters
(i.e., leaves) alphabetically within each depth, as described in the previous
section. Let $S[1,n]$ be the sequence of depths in the canonical code-tree, so
that $S[c]$ ($1 \le c \le n$) is the depth of the character $c$. Now, the depth
and offset of any $c \in \Sigma$ is easily computed from the wavelet tree of
$S$: the depth is just $S[c]$,
while the offset is $rank_{S[c]}(S,c)$. Inversely, given a depth $d$ and an
offset $o$, the corresponding character is $select_d(S,o)$. The $\Oh{w^2}$-bit
data structure of Gagie and Nekrich~\cite{GN09} converts in constant time
pairs {\em (depth,offset)} into codes and vice versa, whereas the multiary
wavelet tree on $S$ requires $n\log L + \Oh{\frac{n\log\log n}{\log_L n}}$ bits
of space and completes encoding/decoding in time
$\Oh{\frac{\log L}{\log\log n}}$. Under the restriction $1/\epsilon =
\polylog{n}$, the space is $\Oh{w^2} + n\log L + o(n)$ and the time is $\Oh{1}$.

Going further, we note that \(H_0 (S)\) is at most \(\log (L - \lceil \log n
\rceil + 1) + \Oh{1}\), so we can store $S$ in $\Oh{n \log (L - \log n + 1) +
n} = \Oh{n\log\log(1/\epsilon)}$ bits. To see this, consider $S$ as two
interleaved subsequences, $S_1$ and $S_2$, of length $n_1$ and $n_2$,
with $S_1$ containing those lengths less than or equal to \(\lceil \log n
\rceil\) and $S_2$ containing those greater. Thus $nH_0(S) \le n_1H_0(S_1) +
n_2H_0(S_2) + n$.

Since there are at most $2^\ell$ codewords of length $\ell$, assume we complete
$S_1$ with spurious symbols so that it has exactly $2^\ell$ occurrences of
symbol $\ell$. This completion cannot decrease $n_1H_0(S_1)
= \sum_{1 \le c \le \lceil\log n\rceil} n_c \log \frac{n_1}{n_c}$, as
increasing some $n_c$ to $n_c+1$ produces a difference of $f(n)-f(n_c)\ge 0$,
where $f(x) = (x+1)\log(x+1)-x\log x$ is increasing. Hence we can reason as if
$S_1$ contained exactly $2^\ell$ occurrences of symbol $1 \le \ell \le \lceil
\log n \rceil$, where
straightforward calculation shows that \(n_1 H_0 (S_1) = \Oh{n_1}\).

On the other hand, $S_2$ contains at most \(L - \lceil \log n \rceil\)
distinct values, so \(H_0 (S_2) \leq \log (L - \lceil \log n \rceil)\), unless
\(L = \lceil \log n \rceil\), in which case $S_2$ is empty and \(n_2 H_0 (S_2)
= 0\).  Thus $n_2 H_0(S_2) \le n_2 \log \lceil \log(2/\epsilon)\rceil =
\Oh{n_2 \log\log (1/\epsilon)}$.

Combining both bounds, we get $H_0(S) = \Oh{1+\log\log (1/\epsilon)}$.
If we assume the text to encode is of length $m = 2^{\mathcal{O}(w)}$, as
usual under the RAM model of computation, then $L = \Oh{w}$ and the following
theorem holds.

\begin{theorem} \label{thm:additive}
For any $0<\epsilon<1/2$ with \(1 / \epsilon = \Oh{\polylog{n}}\), and under
the RAM model with computer word size $w$, so that the text to encode is of
length $2^{\mathcal{O}(w)}$, we can store a prefix code with expected codeword
length within an additive term $\epsilon$ of the minimum, using
$\Oh{w^2 + n \log \log (1 / \epsilon)}$ bits, such that encoding and decoding
any character takes $\Oh{1}$ time.
\end{theorem}

In other words, under mild assumptions, we can store a code using
$\Oh{n\log\log(1/\epsilon)}$ bits at the price of increasing the average
codeword length by $\epsilon$, and in addition have constant-time encoding and
decoding. For constant $\epsilon$, this means that the code uses just $\Oh{n}$
bits at the price of an arbitrarily small constant additive penalty over the
shortest possible prefix code.

\section{Multiplicative increase in expected codeword length} \label{sec:multiplicative}

In this section we focus on a multiplicative rather than additive penalty over
the optimal prefix code, in order to achieve a sublinear-sized representation
of the encoding, which still enables constant-time encoding and decoding.

Our main idea is to divide the alphabet into probable and
improbable characters and to store information about only the probable ones.
Given a constant \(c > 1\), we use Milidi\'u and Laber's algorithm \cite{ML01}
to build a prefix code with maximum codeword length \(L = \lceil \log n \rceil + \lceil 1 / (c - 1) \rceil + 1\).
We call a character's codeword {\em short} if it
has length at most \(L / c + 2\), and {\em long} otherwise.  Notice there are at
most \(2^{L / c + 3} - 1 = \Oh{n^{1 / c}}\) characters with short codewords.
Also, although applying Milidi\'u and Laber's algorithm may cause some
exceptions, characters with short codewords are usually more probable than
characters with long codewords. We will hereafter call {\em infrequent
characters} those encoded with long codewords
in the code of Milidi\'u and Laber.

We transform this length-restricted prefix code as described in
Section~\ref{sec:related}, namely, we sort the characters lexicographically
within each depth. We use a dictionary data structure
$F$ due to Fredman, Koml\'os and Szemer\'edi~\cite{FKS84} to store the indices
of the characters with short codewords. This data structure takes $\Oh{n^{1 /
c} \log n}$ bits and supports membership queries in $\Oh{1}$ time, with
successful queries returning the target character's codeword. We also build
\(\lfloor L / c \rfloor + 2\) arrays that together store the indices of all
the characters with short codewords; for \(1 \leq \ell \leq \lfloor L / c
\rfloor + 2\), the $\ell$th array stores the indices the characters with
codewords of length $\ell$, in lexicographic order by codeword.  Again, we
store the first codeword of each length in $\Oh{w^2}$ bits overall,
following Gagie and Nekrich \cite{GN09}, such that it takes $\Oh{1}$ time to
compute any codeword given its length and offset, and vice versa.  With these
data structures, we can encode and decode any character with a short codeword
in $\Oh{1}$ time.  To encode, we perform a membership query on the dictionary
to check whether the character has a short codeword; if it does, we receive the codeword itself as satellite information returned by the query.  To decode, we first find the codeword's length $\ell$ and offset $j$ in $\Oh{1}$ time as described in Section~\ref{sec:related}.  Since the codeword is short, \(\ell \leq \lfloor L / c \rfloor + 2\) and the character's index is stored in the $j$th cell of the $\ell$th array.

\begin{figure}[t]
\begin{center}
\resizebox{\textwidth}{!}
{\rotatebox{-90}
{{\includegraphics{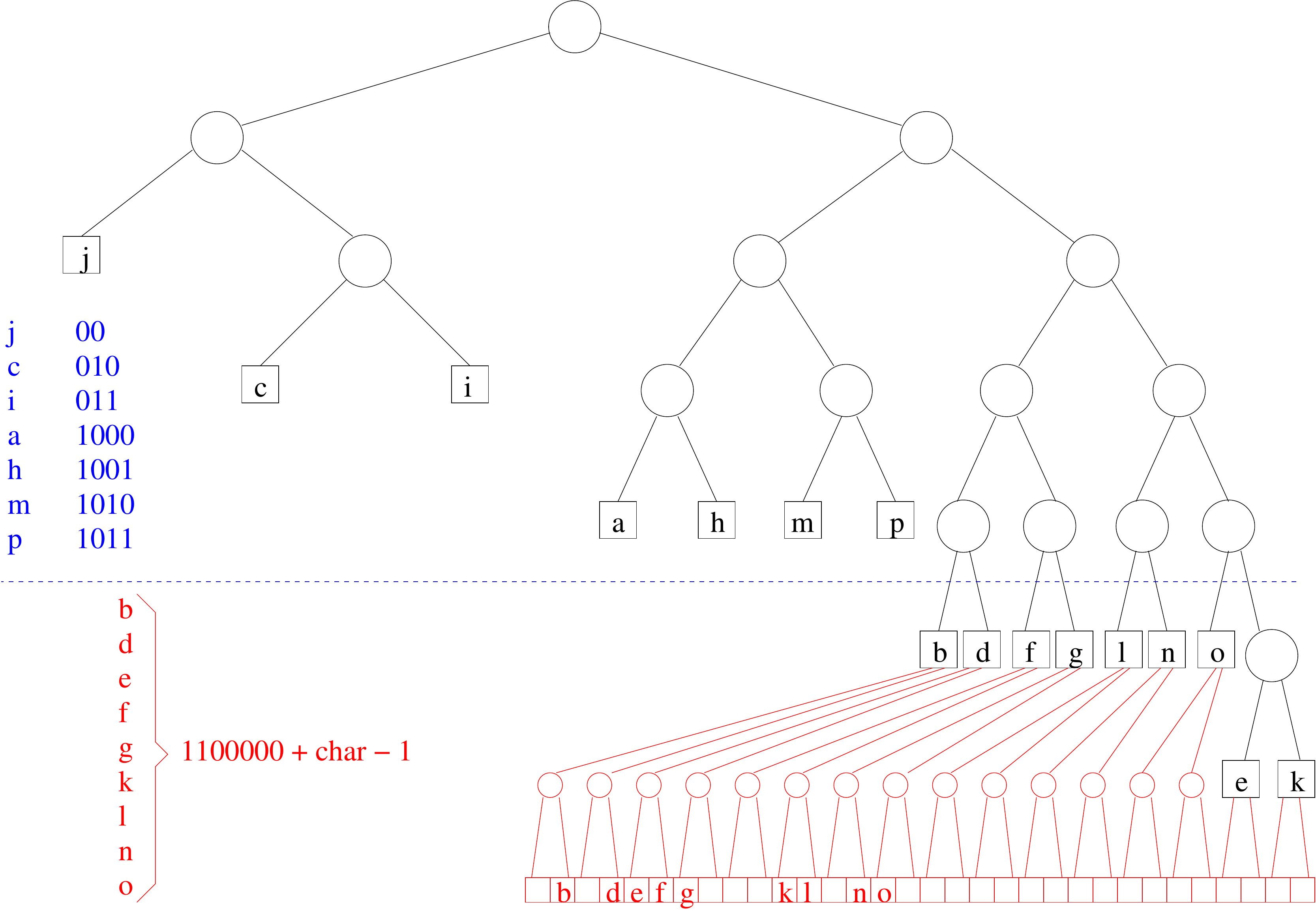}}}}
\end{center}
\caption{An example with $n=16$ and $c=3$. The black tree is the result of applying the algorithm of Milidi\'u and Laber on the original prefix code. Now, we set $L=6$ according to our formula, and declare short the codeword lengths up to $\lfloor L/c \rfloor+2 = 4$. Short codewords are stored unaltered in a dictionary (blue). Longer codewords are changed: All are extended up to length $L+1=7$ and reassigned a code according to their values in the contiguous slots of length 7 (in red).}
\label{fig:thm2}
\end{figure}

We replace each long codeword with \emph{new} codewords: instead of
a long codeword $\alpha$ of length $\ell$, we insert $2^{L+1-\ell}$ new codewords
$\alpha\cdot s$, where $\cdot$ denotes concatenation and $s$ is an arbitrary binary
string of length $L+1- \ell$.  Figure~\ref{fig:thm2} shows an example.  Since \(c > 1\), we have \(n^{1 / c} < n / 2\) for sufficiently large $n$, so we can assume without loss of generality that there are fewer than $n/2$
short codewords; hence, the number of long codewords is at least $n/2$.
  Since every long codeword is replaced by at
least two new codewords, the total number of new codewords is at least $n$.
Since new codewords are obtained by extending all codewords of length
$\ell > L/c+1$ in a canonical code, all new codewords are binary representations
of consecutive integers. Therefore the $i$-th new codeword equals to
$\alpha_f+i-1$, where $\alpha_f$ is the first new codeword.
If $a$ is an infrequent character, we encode it with the $a$-th new codeword,
$\alpha_f+a-1$.
To encode a character $a$, we check whether $a$ belongs to the dictionary $F$.
If $a\in F$, then we output the codeword for $a$. Otherwise we encode $a$
as $\alpha_f+a-1$. To decode a codeword $\alpha$, we read its prefix bitstring
$s_\alpha$ of length $L+1$ and compare $s_\alpha$ with $\alpha_f$. If
$s_\alpha \geq \alpha_f$, then $\alpha=s_\alpha$ is the codeword for
$s_\alpha-\alpha_f +1$. Otherwise, the codeword length
of the next codeword $\alpha$ is at most $L/c+1$ and $\alpha$ can be decoded as
described in the previous paragraph.

By analysis of the algorithm by Milidi\'u and Laber~\cite{ML01} we can see
that the codeword length of a character in their length-restricted code
exceeds the codeword length of the same character in an optimal code by at
most 1, and only when the codeword length in the optimal code is at least \(L - \lceil \log n \rceil - 1 = \lceil 1 / (c - 1) \rceil\).  Hence, the codeword length of a character encoded with a short codeword
exceeds the codeword length of the same character in an optimal code by a factor of at most \(\frac{\lceil 1 / (c - 1) \rceil + 1}{\lceil 1 / (c - 1) \rceil} \leq c\).
Every infrequent character is encoded with a codeword of length $L+1$.
Since the codeword length of an infrequent character in the length-restricted
 code is more than $L/c + 2$, its length in an optimal code is more than $L/c +1$.
Hence,  the codeword length of a long character in our code is at most
 $\frac{L+1}{L/c+1}< c$ times greater than the codeword length of the same
character in an optimal code. Hence, the average codeword length for our code
is  less than $c$ times the optimal one.

\begin{theorem} \label{thm:multiplicative}
For any constant \(c > 1\), under the RAM model with computer word size $w$,
so that the text to encode is of length $2^{\mathcal{O}(w)}$, we can store a
prefix code with expected codeword length within $c$ times the minimum in
$\Oh{w^2+n^{1 / c} \log n}$ bits, such that encoding and decoding any character
takes $\Oh{1}$ time.
\end{theorem}

Again, under mild assumptions, this means that we can store a code with
expected length within $c$ times of the optimum, in $\Oh{n^{1/c}\log n}$ bits
and allowing constant-time encoding and decoding.




\bibliographystyle{psc}
\bibliography{psc09}

\end{document}